\documentclass[twocolumn]{aastex631}

\usepackage{upgreek}

\newcommand{\HI}{H\,\textsc{i}\,}

\shorttitle{$B-\rho $ scaling}
\shortauthors{Auddy, Basu, \& Kudoh}
\graphicspath{{./}{figures/}}

\begin{document}

\title{The Magnetic Field versus Density relation in Star-Forming Molecular Clouds}

\author[0000-0003-3784-8913]{Sayantan Auddy}
\affiliation{Department of Physics and Astronomy, Iowa State University, Ames, IA, 50010, USA
}
\email{sauddy@iastate.edu, sayantanauddy21@gmail.com}
\author[0000-0003-0855-350X]{Shantanu Basu}
\affiliation{Department of Physics and Astronomy, The University of Western Ontario, London, ON N6A 3K7, Canada 
}
\email{basu@uwo.ca}

\author[0000-0003-1292-940X]{Takahiro Kudoh}
\affiliation{Faculty of Education, Nagasaki University, 1-14 Bunkyo-machi, Nagasaki 852-8521, Japan}







\begin{abstract}
We study the magnetic field to density ($B-\rho$) relation in turbulent molecular clouds with dynamically important magnetic fields using nonideal three-dimensional magnetohydrodynamic simulations. Our simulations show that there is a distinguishable break density $\rho_{\rm T}$ between the relatively flat low density regime and a power-law regime at higher densities. We present an analytic theory for $\rho_{\rm T}$ based on the interplay of the magnetic field, turbulence, and gravity. The break density $\rho_{\rm T}$ scales with the strength of the initial Alfv\'en Mach number $\mathcal{M}_{\rm A0}$ for sub-Alfv\'enic ( $\mathcal{M}_{\rm A0}<1$) and trans-Alfv\'enic ($\mathcal{M}_{\rm A0} \sim 1$)  clouds. We fit the variation of  $\rho_{\rm T}$ for model clouds as a function of $\mathcal{M}_{\rm A0}$, set by different values of initial sonic Mach number $\mathcal{M_{\rm 0}}$ and the initial ratio of gas pressure to magnetic pressure $\beta_{\rm 0}$. This implies that $\rho_{\rm T}$, which denotes the transition in mass-to-flux ratio from the subcritical to supercritical regime, is set by the initial turbulent compression of the molecular cloud. 
\end{abstract}



\section{Introduction} \label{sec:intro}
The criticality of magnetic fields in regulating the process of star formation has been debated for decades. While many theoretical studies \citep[e.g., see reviews by][and references therein]{mou78,shu87,shu99,mou99,wur18} have convincingly demonstrated that the magnetic field is indispensable in the formation of stars, its relative importance over turbulence has emerged only recently from observations. The polarized thermal emission of Galactic dust detected by the Planck satellite \citep{pla16} has shown a clear correlation of relative orientation of the ambient magnetic field and the increasing gas column density $N_{\rm H}$. The less dense structures tend to be aligned with the magnetic field while the orientation becomes perpendicular to the elongations in column density maps when $N_{\rm H} \gtrsim10^{22}$ cm$^{-2}$. This suggests that the magnetic field is dynamically important for the formation of density structures on physical scales ranging from approximately $1-10$ pc.
The estimated field strength using the Davis–Chandrasekhar–Fermi (DCF) method \citep{davis51,cha53} also indicates a strong magnetic field as the corresponding mass-to-flux ratio is subcritical (see Table D.1 in \cite{pla16}) and the turbulence is sub- or trans-Alfv\'enic.

The high $N_{\rm H}$ region, with perpendicular magnetic field, maps to the power-law tail in the column density probability density function (NPDF) \citep{sol19}, implying that the magnetic field is crucial in regulating the gravitational collapse. This is consistent with numerical studies \cite[][using nonideal MHD simulations]{aud18,aud19}, which show that the power-law tail at high column density is an imprint of gravitational contraction due to ambipolar diffusion in strongly magnetized clouds.

Theoretically, the scaling of the magnetic field strength $B$ with gas density $\rho$ has implications on how magnetic fields regulate cloud contraction. Zeeman line splitting observations from low-density \HI and higher density molecular clouds (in OH and CN) compiled by \cite{cru10} revealed at least two distinct characteristics in the $B-\rho$ plot that are separated at the break density $\rho_{\rm T}$. These are a flat part ($B\propto \rho^{0}$) at the low densities ($\rho < \rho_{\rm T}$) and a power-law ($B\propto \rho^{\kappa}$) scaling at high density ($\rho \ge \rho_{\rm T}$).

In spite of several studies, the inferred value of $\kappa$ from both observations and numerical models is mostly uncertain. 

The reported values from Zeeman observations span a domain from $\kappa =2/3$ \citep{cru10} that can be interpreted as a result of isotropic contraction of weakly magnetized flux-frozen gas to $\kappa =1/2$ \citep[][]{tri15} that represents anisotropic flux-frozen collapse with a dynamically important magnetic field. 
Compilations of DCF data by \citet{mye21} and \citet{liu21} find best fits $\kappa = 0.66$ and $0.57$, respectively. Both the Zeeman and DCF samples represent an ensemble of objects of different masses and velocity dispersions, therefore cannot be considered an evolutionary sequence.

The effect of ambipolar diffusion is to somewhat reduce $\kappa$ from its flux-frozen value, and \citet{das21} find that the lifetimes of dense cores in the density range $10^4-10^6$ cm$^{-3}$ has a best fit model of ambipolar diffusion with $\kappa = 0.43$. The measured $\kappa$ values from numerical simulations of magnetized self-gravitating clouds are equally varied \citep[see for example][]{kud07,col11,col12,li15b,mocz17}.

In this Letter, we refrain from the $\kappa$ debate and focus on the physical origin of the break density $\rho_{\rm T}$ by studying initially subcritical clouds with sub- or trans-Alfv\'enic turbulence. This is primarily because the observed $\kappa$ cannot be directly compared to theoretical/numerical models \citep{li21}. The $B-\rho$ relation inferred from MHD simulations are ``temporal" in origin, as they are inferred from the time evolution of one cloud, compared to Zeeman observations which have ``spatial" information based on collection of clouds at various locations observed at a similar time. Thus, we instead explore how the turbulent compression of a magnetic cloud sets $\rho_{\rm T}$, which denotes the transition from the subcritical to supercritical regime. We investigate the scenario of a strong magnetic field with turbulent compression acting primarily perpendicular to the
magnetic field. Motions along the magnetic field direction can cause an increase in density while $B$ remains unchanged i.e., $B\propto \rho^{0}$, however neutral-ion slip across field lines can also accomplish this. In practice there may also be some partial flux freezing during an initial phase when the gas oscillates about an approximate force-balanced state, in which case $B$ increases weakly as the density increases. This oscillatory phase ends when ambipolar diffusion causes enough neutrals to slip past the field lines to form local supercritical pockets that subsequently evolve with $B\propto \rho^{\kappa}$.
We present a self-consistent theory of the origin of $\rho_{\rm T}$ and test it against a suite of simulations with different initial conditions. We find a direct connection between $\rho_{\rm T}$ and the strength of Alfv\'enic turbulence. 

In Section \ref{sec:Numsim}, we present numerical simulations that study the properties of the $B-\rho$ relation. 
In Section \ref{sec:analytic} we derive an analytic expression for $\rho_{\rm T}$ based on a model of turbulent compression of a magnetized cloud, and compare with the numerical results. In Section \ref{sec:discussion} we discuss and summarize our results.

\section{Magnetic field - density scaling}\label{sec:Numsim}
We investigate the collapse of magnetized turbulent molecular clouds due to gravitationally-driven ambipolar diffusion. The relative strength of gravity and the magnetic field is measured using the mass-to-flux ratio $M/\Phi$, while the  Alfv\'en Mach number $\mathcal{M}_{\rm A0}$ quantifies the relative importance of turbulence and the magnetic field. There exists a critical mass-to-flux-ratio, $(M/\Phi)_{\rm {crit}}$, such that for strong enough magnetic field, i.e., if $M/\Phi < (M/\Phi)_{\rm {crit}}$, the cloud is subcritical and stable against fragmentation as magnetic flux freezing prevents further collapse. If $M/\Phi > (M/\Phi)_{\rm {crit}}$, the cloud is supercritical and is prone to collapse as gravity dominates. Super-Alfv\'enic motions ($\mathcal{M}_{\rm A0}\equiv v_{\rm {t0}}/v_{\rm{A0}} >1 $) signify the dominance of turbulence and sub-Alfv\'enic ($\mathcal{M}_{\rm A0}< 1 $) motions signify the dominance of magnetic pressure.





\begin{figure*}[ht!]
\centering
\includegraphics[scale=0.9]{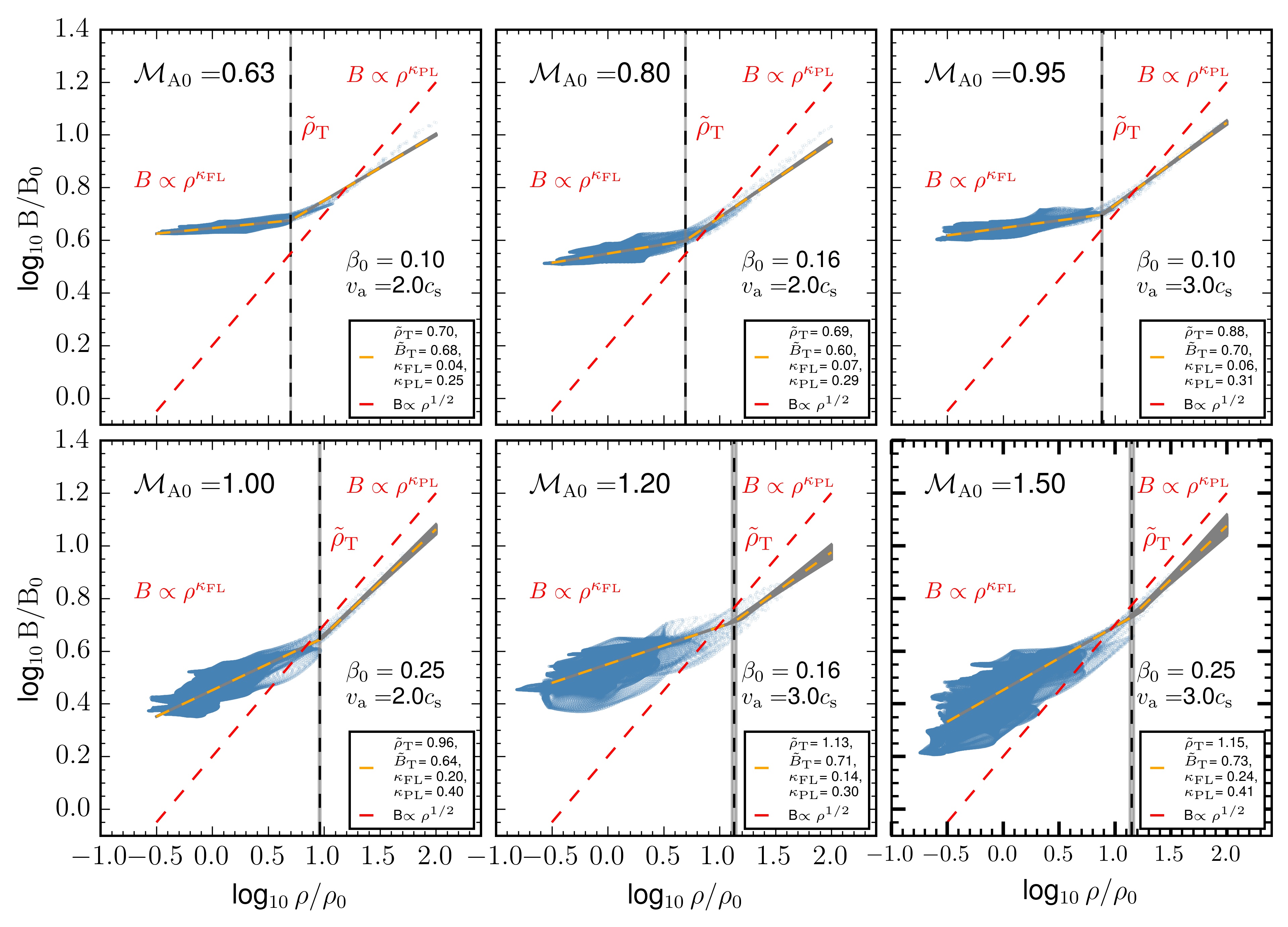}
\caption{Magnetic field strength versus density scaling for simulated models of molecular clouds with different initial Alfv\'en Mach number ($\mathcal{M}_{\rm A0}$) initialized by different values of $\beta_{\rm 0}$ and $v_{\rm a}$ shown in each plot. Each snapshot shows the distribution (in blue) of the midplane values of $B$ and $\rho$  taken at the end of the simulation, when the maximum density has reached $100 \rho_{\rm 0}$. The yellow dashed line represents the mean value from the fit parameters obtained using the MCMC Bayesian fit. The vertical dashed line shows the mean transition density where the flat $B-\rho$ scaling changes to the power-law scaling.  The grey shaded regions represent the standard deviation in the fit parameters.  An animation of the time evolution of the $B-\rho$ scaling in the top middle panel is included as online supplementary material.}\label{fig:B-rho}
\end{figure*}

\subsection{Numerical Model} \label{sec:simulation}
The numerical setup is similar to both \cite{aud18} and \cite{aud19}.
We model clouds with dynamically important magnetic field strength, i.e., sub-critical initial mass-to-flux ratio, under the influence of both sub- and trans-Alfv\'enic turbulence. The clouds are initially stratified in the $z$-direction with a uniform density in the $x-y$ plane and a uniform magnetic field along the $z$-direction: $B_{z}=B_{\rm 0}$, $B_{x}=B_{y} =0$, where $B_{\rm 0}$ is constant. In our model the cloud has already settled to an equilibrium along the $z$-direction, thus we do not model an earlier phase of cloud formation or possibly continuing flow along the field lines \cite[e.g.,][]{vaz11}. We input Gaussian random velocity perturbations in the $x$- and $y$-components with amplitude $v_{\rm a}$ to mimic turbulence using a Fourier spectrum $v_{k}^2 \propto k^{-4}$. The turbulence is allowed to decay and is not replenished during the successive evolution. The number of grid points in each direction are $(N_x, N_y, N_z) = (512, 512, 20)$. The computational domain is $-4\pi H_{\rm 0} \leq x,y \leq 4\pi H_{\rm 0}$ and $0 \leq z \leq 4 H_{\rm 0}$, where  $H_{\rm 0} = c_{\rm {s0}}/\sqrt{2\pi G\rho_{0}}$.



\subsection{Numerical parameters} \label{sec:parameters}
The strength of turbulence is specified by the amplitude $v_{\rm a}$ of the initial velocity perturbation. The initial magnetic field is parametrized using a dimensionless parameter 
\begin{equation}\label{beta}
    \beta_{0} \equiv \frac{8 \pi c_{\rm s0}^{2} \rho_{0}}{B_{0}^{2}},
\end{equation}
which is the ratio of the initial gas to magnetic pressure at $z= 0$. The units of length, velocity and density are set by scale length $H_{\rm 0} = c_{\rm {s0}}/\sqrt{2\pi G\rho_{0}}$, the isothermal sound speed $c_{\rm {s0}}$, and the midplane density $\rho_{\rm 0}$. The dimensional values of all other quantities are obtained by specifying appropriate values for $\rho_{\rm 0}$ and $c_{\rm {s0}}$. For example $H_{0} \simeq 0.05 $ pc and $t_{0} \simeq 2.5 \times 10^{5}\,\rm {yr}$ if $c_{\rm s0} = 0.2$ km s$^{-1}$ and $n_{0} \equiv \rho_{0}/m_n = 10^{4}$ cm$^{-3}$ where $m_n = 2.33 \times 1.67 \times 10^{-24}$ g. Equation (\ref{beta}) yields $B_{0} \simeq$ 50 $\upmu \rm G$ for $\beta_{0} = 0.16$. The initial number column density is $N_0 \equiv \Sigma_{0}/m_{n} \simeq 1.5 \times 10^{21}$ cm$^{-2}$ when the initial column density is $\Sigma_{0} = \rho_{0} H_{0} \simeq 6 \times 10^{-3}$ g cm$^{-2}$.


\subsection{$B-\rho$ scaling} \label{sec:parameters}
The $B-\rho$ scaling can be identified by a relatively flat part at low density and a power law at high density. 
To characterize it we consider a simple piece-wise function
\begin{eqnarray}
\tilde {B} =& \kappa_{\rm FL} \tilde{\rho} + \tilde{B}_{\rm T} - \kappa_{\rm FL} \tilde{\rho}_{T} ,&\, \, \tilde{\rho} < \tilde{\rho}_{\rm T}   \nonumber \\
      =& \kappa_{\rm PL} \tilde{ \rho }+  \tilde{ B}_{\rm T} - \kappa_{\rm PL} \tilde{ \rho}_{T}  ,&\, \, \tilde{\rho} \ge \tilde{\rho}_{\rm T},
\end{eqnarray}
where $\tilde {B}\equiv \log B/B_{\rm 0}$, $\tilde{\rho}\equiv \log \rho/\rho_{\rm 0}$, $\kappa_{\rm FL}$ and  $\kappa_{\rm PL}$ are the slopes of the flat and the power-law region, respectively. The break density $\tilde{\rho_{\rm T}} = \log \rho_{T}/\rho_{\rm 0}$ and the corresponding magnetic field strength $\tilde {B_{\rm T}} = \log B_{\rm T}/B_{\rm 0}$ mark the transition from the flat low density part to the power-law high density regime. We fit the four free parameters using the Markov Chain Monte Carlo (MCMC) method \citep{van03}. We use the \textsc{python} package {\it emcee} \citep{mac13} for this purpose. 

Figure \ref{fig:B-rho} shows the $B-\rho$ relation for six different models with different Alfv\'en Mach number ($\mathcal{M}_{\rm A0}$), summarized in Table \ref{tab:summary},  initialized by different choices of $\beta_{\rm 0}$ and sonic Mach number $\mathcal{M}_{\rm 0} \equiv \sqrt{2}v_{\rm a}/c_s$. Note that $\mathcal{M}^2_{\rm 0} \beta_{0} \equiv 2 \mathcal{M}^2_{\rm A0} $. All the models initially have subcritical mass-to-flux ratio, with collapse mainly regulated by the turbulence accelerated ambipolar diffusion. Each snapshot consists of the midplane values $B$ and $\rho$ taken at the end of the simulation when the maximum density has reached $100 \rho_{\rm 0}$. The best-fitting parameters  $\kappa_{\rm FL}$, $\kappa_{\rm PL}$, $\tilde{\rho}_{\rm T}$, and $\tilde{B}_{\rm T}$ are shown in each of the panels. The plots are arranged from the top left to the bottom right according to the increasing values of the  Alfv\'en Mach number ($\mathcal{M}_{\rm A0}$). The yellow dashed line represents the mean value of the fit parameters obtained using the MCMC Bayesian fit. The vertical dashed line shows the mean transitional density where the flat $B-\rho$ scaling changes to the power-law scaling.  The grey shaded regions represent the standard deviation in the fit parameters. 

There are at least two distinct characteristics that emerge from Figure \ref{fig:B-rho}. First, the break density $\rho_{\rm T}$ shifts towards higher density with increasing values of $\mathcal{M}_{\rm A0}$. For example, it is minimum for model T210 ($\mathcal{M}_{\rm A0} =0.63 $) and maximum for  model T325 ($\mathcal{M}_{\rm A0} =1.50 $) with $\rho_{\rm T}/\rho_{\rm 0}= 5.01$ and $\rho_{\rm T}/\rho_{\rm 0}= 14.13$ respectively. For molecular clouds with mean number density $n_{\rm 0} \approx 10^{2}$ cm$^{-3}$ the transition density $n_{\rm T} = (\rho_{\rm T} / \rho_{\rm 0}) n_{\rm 0}$ corresponds to $\approx 500 - 1400$ cm$^{-3}$ for the range of $\rho_{\rm T}/\rho_{\rm 0}$. Second, the low density region ($\rho < \rho_{\rm T}$) gets less flat and more scattered with an increasing strength of $\mathcal{M}_{\rm A0}$. 

\begin{figure}
\centering
\includegraphics[scale=.80,]{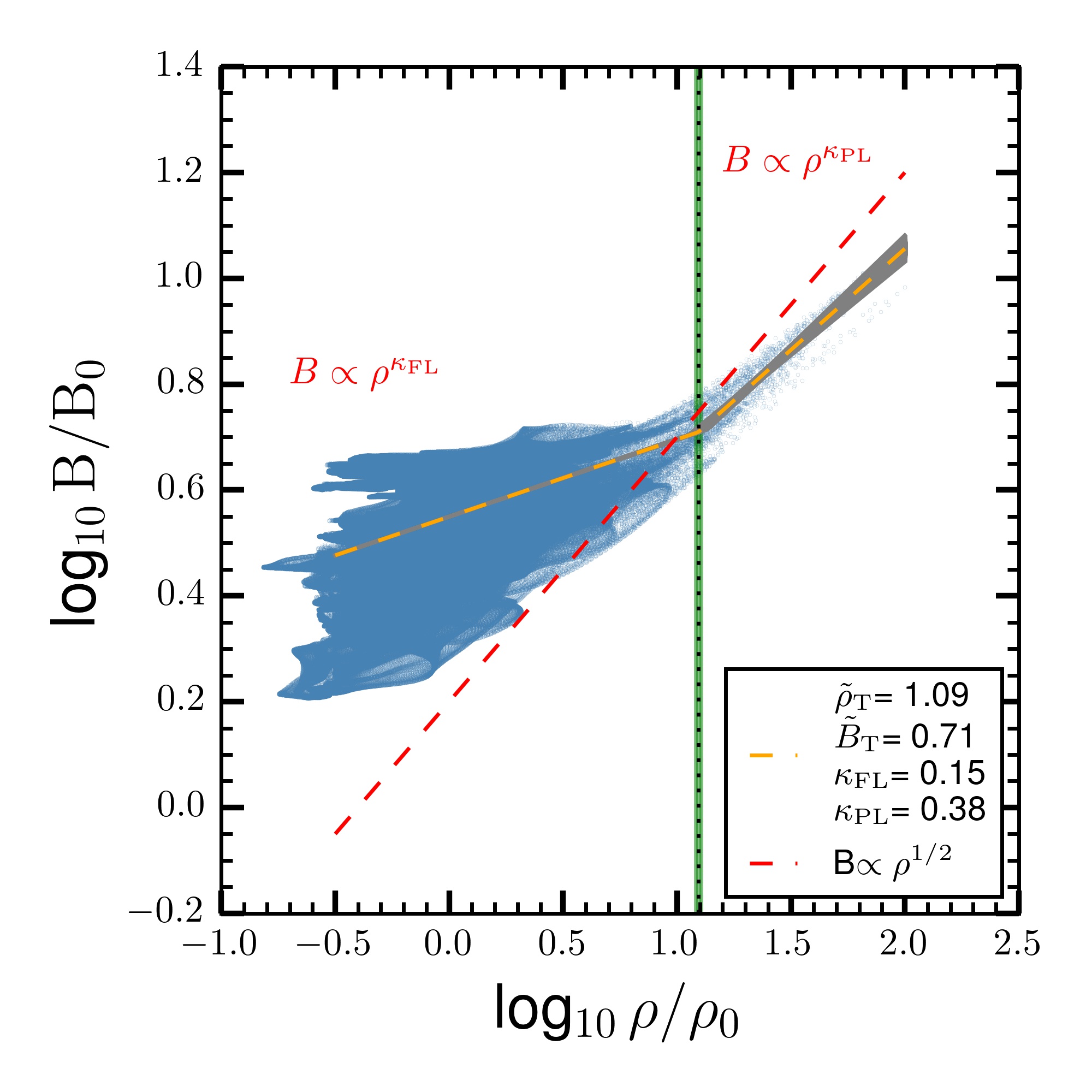}
\caption{Magnetic field strength versus density scaling from all the simulations put together.}\label{completeFigure}
\end{figure}

Figure \ref{completeFigure} represents the composite data from all the simulations explored above. It implies that observations of magnetic fields from an ensemble of clouds that have different initial conditions will yield a much larger scatter in the low density regime than in the high density regime. The latter converges to a narrower range of collapsing regions with similar values of their mass-to-flux ratio and less effect of ambipolar diffusion.





\begin{table}
	\centering
	\caption{Model and Fit Parameters}
	\label{Parameter}
	\begin{tabular}{lllllc} 
		\hline
		\hline
		Model & $v_{\rm {a}} / c_{\rm {s}}$  & $\beta_0$ & $\mathcal{M}_{\rm A0}$ & Comments  \\
		\hline
		T210 &  2.0  & 0.10   & 0.63  & sub-Alfv\'enic  \\
		T216 &  2.0  & 0.16   & 0.80  & sub-Alfv\'enic  \\
		T310 &  3.0  & 0.10   & 0.95  & sub-Alfv\'enic  \\
		T225 &  2.0  & 0.25   & 1.00  & trans-Alfv\'enic\\
		T316 &  3.0  & 0.16   & 1.20  & trans-Alfv\'enic\\
		T325 &  3.0  & 0.25   & 1.50  & super-Alfv\'enic\\

		\hline
	\end{tabular}
	\tablecomments{ $\beta_{0} $ is the initial ratio of thermal to magnetic pressure at $z = 0$, $v_{\rm a}$ is the amplitude of the initial velocity fluctuation, 
	and $\mathcal{M}_{\rm A0}$ is the initial Alfv{\'e}n Mach number.}\label{tab:summary}
\end{table}

\section{Analytic Model}\label{sec:analytic}
The physical origin of a transition density in clouds with dynamically important magnetic field
can be explained using an analytic model. The cloud flattens along the ambient magnetic field and the subsequent evolution is primarily perpendicular to the magnetic field lines. The turbulence compression acts primarily  perpendicular to the magnetic field direction and establishes a local pressure balance between the ram and magnetic pressure. This leads to the formation of magnetic ribbons \citep{kud14,aud16,aud18} till gravitationally-driven ambipolar diffusion creates supercritical pockets of dense regions that are prone to collapse.

We consider a local pressure balance in a cloud stratified along the $z$-direction with compression along the $x-y$ plane. For simplicity, we assume that the thermal pressure is negligible compared to the magnetic and ram pressure.  On compression the magnetic strength $B$ increases until the magnetic pressure in the compressed region balances the external ram pressure and the initial pressure due to the background magnetic field  $B_{\rm 0}$. This results in a quasi-equilibrium state where

\begin{equation}\label{forcebalance}
H\,\frac{B^2}{8 \pi} = H_{0}\left(\rho_{0}v_{\rm {t0}}^{2} + \frac{B_{0}^2}{8 \pi}\right),
\end{equation}
and $v_{\rm {t0}} \equiv \sqrt{2}v_{\rm a}$ is the nonlinear flow speed. The compression slowly ceases and oscillations ensue. The gas establishes a hydrostatic equilibrium as it settles along the $z$-direction such that the cloud has a half-thickness 

\begin{equation}\label{scaleheight}
H= \frac{c_{\rm s}}{\sqrt{2 \pi G \rho}}
\end{equation}
\citep{spi42}. If the ambipolar diffusion time scale is much longer than the compression time, the cloud is nearly flux frozen during the initial compression, i.e., $B/\Sigma = B_{\rm 0}/\Sigma_{\rm 0}$. For the column density $\Sigma = 2\rho H$ the flux-frozen relation can be rewritten as 

\begin{equation}\label{fluxfreezing}
    \frac{B}{\rho^{1/2}} =\frac{B_{0}}{\rho_{\rm 0}^{1/2}}\, .
\end{equation}

Simplifying  Equation (\ref{forcebalance}) using Equations (\ref{scaleheight}) and (\ref{fluxfreezing})  we get 

\begin{equation}\label{FB2} 
\left(\frac{\rho}{\rho_{0}}\right)^{1/2} =  v_{\rm {t0}}^2 \left(\frac{8 \pi \rho_{0}}{B_{0}^2} \right) + 1 \,.
\end{equation}
The force balance of Equation (\ref{FB2}) gives the transition density, subsequently denoted as $\rho_{\rm T}$. As the initial turbulence slowly decays and magnetic flux decays due to ambipolar diffusion, the subsequent oscillation slows down. Allowing for such variation we rewrite  Equation (\ref{FB2}) in terms of Alfv\'en Mach number $\mathcal{M}_{\rm A0}\equiv v_{\rm t0}/v_{\rm A0} $ as 
\begin{equation}\label{FB3} 
\left(\frac{\rho_{\rm T}}{\rho_{0}}\right)^{1/2} =a\,( 2 \mathcal{M}_{\rm A0}^2 + 1)\, ,
\end{equation}
where $v_{\rm A0} \equiv B_{0}/ \sqrt{4 \pi \rho_{0}} $ is the initial Alfv{\'e}n speed in the midplane and $a$ is the correction factor that captures the uncertainties about the turbulent decay and flux loss.

\begin{figure}
\begin{center}
	\includegraphics[scale=.60,]{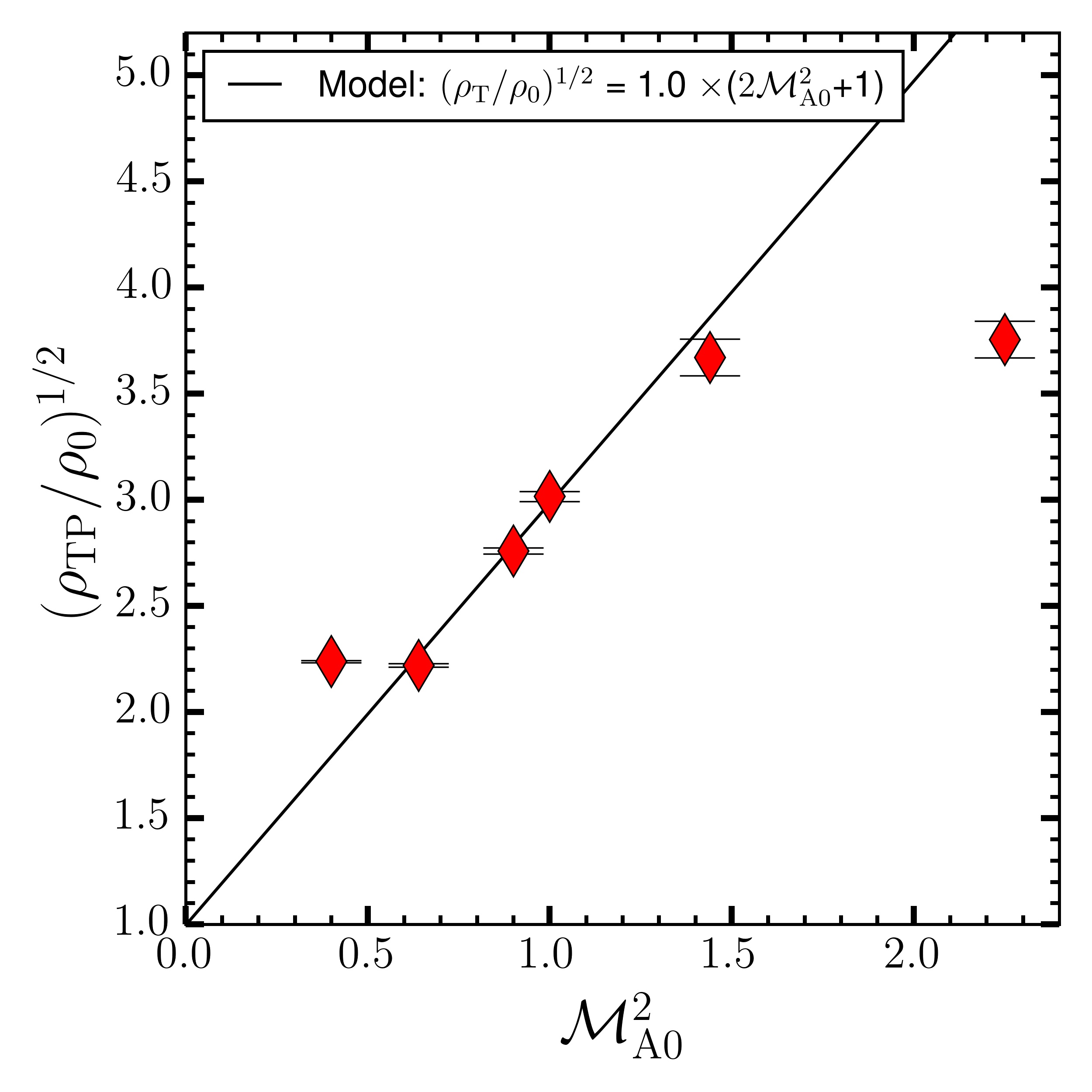}
    \caption{The square-root of the normalized break density $\rho_{\rm T}/\rho_{\rm 0}$ obtained from the simulations for different values of the squared initial Alfv{\'e}n Mach number $\mathcal{M}_{\rm A0}^2$. The black line is the theoretical model of Equation (\ref{FB3}) with the best-fit value $a = 1.0 \pm 0.1$. }
    \label{fig:theoryfit}
\end{center}
	\end{figure}

Figure \ref{fig:theoryfit} shows the variation of the square-root of the normalized break density $(\rho_{\rm T}/\rho_{\rm 0})$ for the simulated models with different values of the square of the Alfv{\'e}n Mach number $\mathcal{M}_{\rm A0}$. There is a good agreement between the simulation data and our analytic Equation (\ref{FB3}) with a best-fit value $a = 1.0 \pm 0.1$. The outlier at  $\mathcal{M}_{\rm A0}^{2} = 2.25$ shows a break in the correlation as $\mathcal{M}_{\rm A0}$ (=1.5) becomes super-Alfv{\'e}nic.
We did not include this point in our fitting since our analytic model breaks down in the super-Alfv{\'e}nic regime, but if we did include it then the best fit is very similar with $a = 0.9$.

\section{Discussion and conclusions}\label{sec:discussion}
We have presented the scaling relation of density and the magnetic field obtained from nonideal three-dimensional MHD simulations. Unlike most other numerical experiments, with ideal MHD, our simulations enable us to investigate gravitational collapse in clouds with much stronger magnetic fields due to the inclusion of ambipolar diffusion.  The relatively flat part (close to $B\propto \rho^{0}$) at the low density regime is a consequence of the subcritical cloud undergoing flattening due to self-gravity along the field lines and settling into a vertical hydrostatic equilibrium, accompanied by some neutral-ion slip. The $B-\rho$ relation in the low density region gets less flat and more scattered with the increasing strength of $\mathcal{M}_{\rm A0}$ as a partial flux freezing becomes more relevant. There is a gradual loss of magnetic flux with each successive oscillation of the compressed regions. The ambipolar diffusion causes the time-averaged force balance between the magnetic field and the ram pressure perpendicular to the magnetic field lines to gradually relax. 
The break density $\rho_{\rm T}$ defines the cutoff beyond which the mass-to-flux ratio becomes supercritical. For $\rho > \rho_{\rm T}$, gravity becomes dominant and the cloud collapses, resulting in $B\propto \rho^{\kappa}$. Our model shows that $\rho_{\rm T}$ is a measure of the initial Alfv\'en Mach number $\mathcal{M}_{\rm A0}$. 

One can identify at least three distinct but related transitions in the structure of molecular clouds (on physical scales spanning $\approx 1-10$ pc) that are signatures of a strong magnetic field. In addition to  $\rho_{\rm T}$, the change of the relative orientation of the ambient magnetic field with increasing gas column density \citep{pla16,sol19} as well as the transition from lognormal to power-law forms in NPDFs at the transitional column density $\Sigma_{\rm TP}$ \citep{aud19} are all imprints of magnetic field regulated star formation. These observations, supported by theoretical models, highlight the interactions of the magnetic field with both gravity and turbulence and supports the paradigm of magnetic field regulated star formation.

We conclude by highlighting three key results:

\begin{itemize}
\item The transition density $\rho_{\rm T}$ separates the turbulent magnetically dominated region, where the $B-\rho$ scaling is mostly flat, from the power-law slope where the cloud becomes supercritical and collapses under gravity

\item With increasing strength of the initial Alfv{\'e}n Mach number $\mathcal{M}_{\rm A0}$, the extent of flattening diminishes as the low density region gets steeper and more scattered

\item 
The transition density $\rho_{\rm T}$ is a function of the Alfv{\'e}n Mach number $\mathcal{M}_{\rm A0}$ and increases with the increasing strength of $\mathcal{M}_{\rm A0}$

\end{itemize}

\begin{acknowledgments}
 We thank the anonymous referee for constructive comments and feedback.Computations were carried out using facilities provided by 
SHARCNET and Compute Canada. S.A. acknowledges support from NASA under {\em Emerging Worlds} through grant 80NSSC20K0702. S.B. is supported by a Discovery Grant from NSERC.
\end{acknowledgments}

\newpage


\bibliography{sample631}{}
\bibliographystyle{aasjournal}



\end{document}